\documentclass[fleqn,12pt,twoside]{article}
\usepackage{espcrc1}
\usepackage{epsfig}

\newcommand{\lsim}{\, \, \raisebox{-0.8ex}{$\stackrel{\textstyle <}{\sim}$ }}
\newcommand{\gsim}{\, \, \raisebox{-0.8ex}{$\stackrel{\textstyle >}{\sim}$ }}

\def\beq{\begin{equation}}
\def\eeq{\end{equation}}
\def\bea{\begin{eqnarray}}
\def\eea{\end{eqnarray}} 
\def\beqa{\begin{equation}\begin{array}{l}}
\def\eeqa{\end{array}\end{equation}}

\newcommand{\AmS}{{\protect\the\textfont2
  A\kern-.1667em\lower.5ex\hbox{M}\kern-.125emS}}

\title{M. Poincar\'e visits Jefferson Lab: Relativistic Models
of Few-Nucleon Systems}

\author{Daniel R.\ Phillips\\
Department of Physics and Astronomy,
Ohio University, Athens, OH 45701%
}
      
\begin{document}

\maketitle

\begin{abstract}
\noindent
{\small I discuss relativistic models of few-nucleon systems, with
particular emphasis on calculations of electron-deuteron scattering
and the comparison of these calculations with recent data from
Jefferson Lab.}
\end{abstract}

\bigskip

\section{Introduction}

In the ``standard model of nuclear physics'' nuclear properties are
calculated using a Schr\"odinger equation in which the degrees of
freedom are nucleons interacting via energy-independent
potentials. This picture can now be solved essentially exactly for
light nuclei~\cite{CS98,Pi03}. However, it appears it only describes
non-relativistic systems, since the Poincar\'e algebra is satisfied
approximately: order-by-order in an expansion in momenta over the
nucleon mass~\cite{Fo95,KF74}.

Here I review attempts to build a phenomenology of light nuclei which
satisfies the Poincar\'e algebra exactly, and so is manifestly
applicable when few-nucleon systems are probed at GeV-scale momentum
transfers---the kinematic domain accessed at Jefferson Lab. I will
focus on elastic electron-deuteron scattering, since it involves the
simplest non-trivial nucleus, and yet also requires the calculation of
$NN$-system wave functions away from the two-body centre-of-mass
frame.

In Section~\ref{sec-rel} I write down the Poincar\'e algebra, and
sketch some of the different means by which model-builders have
attempted to satisfy it. Then in Section~\ref{sec-ed} I describe the
calculation of matrix elements for the interaction of the probe
(electron) with a relativistic bound state (the deuteron). I enumerate
the places in which ``relativistic effects'' can occur in the
calculation, and show one explicit example of such an effect.  I then
compare a number of relativistic models of $e$d scattering, and
address the issue of how we should interpret their widely-varying
predictions. I conclude in Section~\ref{sec-conc}.

\section{What is a relativistic model?}
\label{sec-rel}

The laws of quantum mechanics are the same in all inertial frames of
reference.  In {\it non-relativistic} quantum mechanics the generators
of the transformations relating one frame of reference to another are
the total momentum ${\bf P}$ (translations), the total angular
momentum ${\bf J}$ (rotations), and the boost operator ${\bf K}_G$.
This last operator can be defined by its action on state vectors:
suppose Abigail and Beatrice have their coordinate systems aligned at
$t=0$, but Abigail moves with velocity ${\bf v}$ relative to Beatrice:
\begin{equation}
\langle {\bf x}|\psi(t) \rangle_A \, = \, \langle {\bf x} + {\bf v}
t|\psi(t)\rangle_B \, \equiv \, \langle {\bf x}|e^{i {\bf K}_G \cdot {\bf v}}
|\psi(t) \rangle_B.
\end{equation}

Considering the combined effect on a state vector of all possible
pairs of translations, rotations, and boosts then leads to the
following commutation relations:
\begin{equation}
[P^i,P^j]=0; \qquad [J^i,J^j]=i \epsilon^{ijk} J^k; \qquad
[J^i,K_G^j]=i \epsilon^{ijk} K_G^k;
\label{eq:gal1}
\end{equation}
\begin{equation}
[J^i,P^j]=i \epsilon^{ijk} P^k; \qquad 
[K_G^i,K_G^j]=0;  \qquad [K_G^i,P^j]=-i \delta^{ij} M.
\label{eq:gal2}
\end{equation}
Meanwhile, the invariance of Schr\"odinger equation dynamics means
that the Hamiltonian, $H$, commutes with ${\bf J}$ and ${\bf P}$.  $H$
does {\it not} commute with ${\bf K}_G$:
\begin{equation}
[H,P^i]=0; \qquad [H,J^i]=0; \qquad [H,K_G^i]=i P^i.
\label{eq:gal3}
\end{equation}
The relations (\ref{eq:gal1})--(\ref{eq:gal3}) define the Galilei
algebra, which relates observations in different frames of reference,
and must be obeyed in  non-relativistic quantum mechanics.

The Poincar\'e algebra also relates observations in different frames,
but now it implements Einsteinian, rather than Galilean,
relativity. The changes in the algebra seem simple enough: the only
commutators which change are the last two in (\ref{eq:gal2}).
They become:
\begin{equation}
[K^i,K^j]=-i \epsilon^{ijk} J^k; \qquad [K^i,P^j]=-i \delta^{ij} H.
\label{eq:poinc}
\end{equation}

{\it Any ``relativistic'' quantum theory should obey the Poincar\'e
algebra.} The first of the two relations (\ref{eq:poinc}) states that
relativistic boosts ${\bf K}$, in contrast to their Galilean cousins,
are not commutative. The second is the key commutator which makes
relativistic quantum theories difficult to construct: it encodes the
absence of absolute time in relativistic theories, since it connects
the generator of time translations, $H$, to the generators of spatial
boosts and translations. In any interacting theory $H$ is the sum of a
free part $H^0$ and a potential energy $V$, so Eq.~(\ref{eq:poinc})(b)
means that either ${\bf K}$ or ${\bf P}$ must depend on $V$.

This is one way to understand Dirac's famous ``three forms'' of
relativistic quantum mechanics~\cite{Di49}. In {\it instant form}
(which will be the form I mainly discuss here) ${\bf J}$ and ${\bf P}$
depend on $H^0$ alone, and so are the same in the interacting theory
as they are for free particles. In contrast, ${\bf K}$ depends on $V$,
and so will be different for systems with different dynamics.  We say
that the angular and linear momentum are kinematical generators, while
boosts are ``dynamical''. In {\it point form} ${\bf P}$ is dynamical,
and ${\bf K}$ is kinematic. Finally, in {\it front form} linear
combinations of the 9 Poincar\'e algebra commutators are taken, such
that only two, rather than three, generators need involve $V$. In
front form rotations about the $\hat{x}$ and $\hat{y}$ axes are
dynamical. It has, though, the great advantage that boosts in the
$\hat{z}$-direction are kinematic.

One example of theories that manifestly obey the Poincar\'e algebra
are relativistic quantum field theories (QFT), such as Quantum
Electrodynamics. Field theories can be quantized so that they use
instant-form, front-form, or point-form dynamics~\footnote{See
Ref.~\cite{CM01} for calculations of electron-deuteron scattering in a
front-form QFT-based approach.}. If instant form is adopted then ${\bf
J}$ and ${\bf P}$ can be written down directly from the
stress-energy-momentum tensor of the free theory, and do not connect
different Fock-space sectors of the theory~\footnote{I do not consider
problems of gauge invariance in such definitions here.}. In contrast,
${\bf K}$ inolves the particle-number-changing interaction piece of
the QFT Hamiltonian, and so connects different sectors of the Fock
space.  In a weak-coupling theory perturbative forms of the boost are
useful, but in nuclear physics, we deal with coupling constants of
order 1, and some non-perturbative approximation to the boost, which
manages to preserve---at least approximately---the Poincar\'e
algebra, is needed. This means that the application of QFT to
relativistic nuclear physics is not straightforward.

{\it Manifest} covariance is a sufficient, but not necessary,
condition for a theory to be relativistic. The Bakamjian-Thomas (BT)
construction~\cite{BT53} is a technique by which a ${\bf P}$, ${\bf
J}$, and ${\bf K}$ that obey the Poincar\'e algebra may be
constructed, once a potential energy $V$ is given. Crucially, particle
number is conserved in the BT construction---as long as $V$ does not
change it. In instant form, we first introduce the intrinsic spin
${\bf j}$, the operator ${\bf X}=i \nabla_{\bf P}$, and the invariant
mass of the system $M$. Some algebra then shows that
defining~\cite{Ke94}:
\begin{equation}
H=\sqrt{M^2 + {\bf P}^2}; \qquad {\bf J}={\bf j} + {\bf X} \times
{\bf P}; \qquad {\bf K}=-\frac{1}{2} \{H,{\bf X}\} - \frac{{\bf P}
\times {\bf J}}{H + M},
\end{equation}
gives a set of operators that obeys the Poincar\'e algebra.  Note that
${\bf K}$ depends on $V$, but the BT ${\bf P}$ and ${\bf J}$ are the
same as in the free theory. BT constructions for front form 
and point form 
are also possible~\cite{Ke94}. Some results for $e$d scattering
employing them will be displayed below.

Suppose then that we have a relativistic quantum theory, with dynamics---an
$H^0$ and a $V$---which is God-given in the rest frame of the
many-body quantum system. Considering, for the time being, only one
state in the spectrum, we write:
\begin{equation}
H_{\bf 0} |\psi \rangle_{\bf 0}=E_{\bf 0} |\psi \rangle_{\bf 0}.
\end{equation}
Relativistic covariance imposes constraints on matrix elements
in this theory. For instance, if ${\cal J}$ is an operator
corresponding to an observable which is a Lorentz scalar, then we should 
have:
\begin{equation}
{}_{\bf 0} \langle \psi| {\cal J}_{\bf 0} |\psi \rangle_{\bf 0}
={}_{\bf Q} \langle \psi| {\cal J}_{\bf Q} |\psi \rangle_{\bf Q},
\label{eq:equiv}
\end{equation}
where the subscripts indicate operators and states constructed by
observers in frames in which the system is, respectively, at rest, and
moving with total momentum ${\bf Q}$. Quantum-mechanical equivalences
of the type (\ref{eq:equiv}) imply the presence of unitary
transformations relating wave functions and operators:
\begin{equation}
|\psi \rangle_{\bf Q}=U_{\bf Q} |\psi \rangle_{\bf 0}; \qquad
{\cal J}_{\bf Q}=U_{\bf Q} {\cal J}_{\bf 0} U_{\bf Q}^\dagger.
\end{equation}
$U_{\bf Q}$ is a unitary representation of the Poincar\'e group.  Of
course, in a manifestly covariant theory (\ref{eq:equiv})
is satisfied automatically, as long as $|\psi \rangle_{\bf Q}$
and ${\cal J}_{\bf Q}$ are constructed consistently.

\section{Electron-deuteron as an example}

\label{sec-ed}

\subsection{Observables}

\begin{figure}[htbp]
\centerline{\psfig{figure=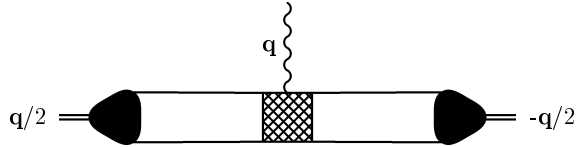, width=16cm}}
\vspace*{-17.8cm}
\caption{Three momenta of the deuteron and virtual photon in the Breit
frame.}
\label{fig-Breit}
\end{figure}

The $e$d differential cross section, can (up to corrections of
$O(\alpha^2)$) be written in terms of two structure functions, $A$ and
$B$:
\begin{equation}
\frac{d \sigma}{d \Omega}=\frac{d \sigma}{d \Omega}_{\rm Mott} \left[A(Q^2) + B(Q^2) \tan^2\left(\frac{\theta_e}{2}\right)\right],
\end{equation}
where $\theta_e$ is the c.m.-frame electron scattering angle,
$q^2=(p_e'-p_e)^2 \equiv -Q^2$ is the virtuality of the photon
exchanged between the electron and the nucleus, and $\frac{d \sigma}{d
\Omega}_{\rm Mott}$ is the Mott $e$d cross-section. $A$ and $B$ are
related to the deuteron form factors $G_C$, $G_Q$, and $G_M$ by:
\begin{equation}
A=G_C^2 + \frac{2}{3} \eta G_M^2 + \frac{8}{9} \eta^2 M_d^4 G_Q^2;
\qquad B=\frac{4}{3} \eta (1 + \eta) G_M^2; \qquad \eta \equiv \sqrt{1 +
\frac{Q^2}{4 M_d^2}}.
\end{equation}
We also consider the tensor-polarization observable, $T_{20}$,
which is defined from the ratios:
\begin{equation}
x=\frac{2}{3} \, \eta \, \frac{G_Q}{G_C}; \qquad
y=\frac{2}{3} \, \eta \left(\frac{G_M}{G_C}\right)^2 
\left[\frac{1}{2} + (1 + \eta) \tan^2 \left(\frac{\theta_e}{2}\right) \right];
\end{equation}
\begin{equation}
T_{20}=\sqrt{2} \frac{x(x + 2) + y/2}{1 + 2 (x^2 + y)}.
\end{equation}
Measurements of $T_{20}$, $A$, and $B$ at a fixed $Q^2$ enable the
extraction of $G_C$, $G_Q$, and $G_M$, as is done, for instance in
Ref.~\cite{Ab00B} out to $Q^2 \sim 2$ GeV$^2$.  In the Breit frame
(see Fig.~\ref{fig-Breit}) the deuteron charge, quadrupole, and
magnetic form factors are linear combinations of matrix elements of
the $NN$ four-current $J^\mu$ between deuteron magnetic substates, for
example:
\begin{equation}
G_C=\frac{1}{3 |e|} \left(\left \langle 1\left|J^0\right|1 \right \rangle + 
\left \langle 0\left|J^0\right|0 \right \rangle + \left \langle -1\left|J^0\right|-1 \right \rangle\right).\label{eq:GC}
\end{equation}

\vspace*{-4.0cm}
\vspace*{-0.3cm}

\subsection{Calculations}

Thus to calculate $A$, $B$, and $T_{20}$ we either need an explicit
calculation of the wave functions for deuteron magnetic substates
$|m_z \rangle_{\pm{\bf q}/2}$, or we
must employ a unitary representation of the Poincar\'e group to
write:
\begin{equation}
{}_{{\bf q}/2}\langle \psi|J_{\rm Breit}^\mu|\psi \rangle_{-{\bf q}/2}
={}_{\bf 0}\langle \psi|U^\dagger_{{\bf q}/2} J_{\rm Breit}^\mu
U_{-{\bf q}/2}|\psi\rangle_{\bf 0}.
\end{equation}
To simplify matters, we now expand $J_{\rm Breit}^\mu=j^\mu_{\rm
Breit} + J^\mu_{\rm 2B, Breit}$, where $j^\mu$ is the single-nucleon
current, and drop $J^\mu_{\rm 2B}$.  This leads to a ``Relativistic
Impulse Approximation'' (RIA).

If, as we are doing here, we take the view that the Poincar\'e group
relates processes in different frames, and remain agnostic about
relativity's impact on $|\psi \rangle_{\bf 0}$, there are two potential
sources of relativistic effects in the RIA. First, the one-body
current:
\begin{equation}
j^\mu_{\rm Breit}=\bar{u}({\bf k}_1') \left[F_1(Q^2) \gamma^\mu
+ F_2(Q^2) i \frac{\sigma^{\mu \nu} q_\nu}{2 M}\right] u({\bf k}_1)
\label{eq:sing}
\end{equation}
contains effects traditionally called ``relativistic'', since they
arise in a Dirac equation treatment of the nucleon, e.g.  Thomas
precession, the Foldy contribution to $\langle r_n^2 \rangle$.

Second, there are boost effects arising from the $U$'s. I will
mention one simple example of these, that the boost 
incorporates length contraction.  In instant form this has the
consequence that the deuteron wave function, computed in the $NN$ rest
frame, is evaluated at a value of ${\bf p}$ which is length-contracted with
respect to the value in the Breit frame~\cite{AA97,Wa01,SP03}:
\begin{equation}
\langle {\bf p}|U_{-{\bf q}/2}|\psi \rangle_{\bf 0} \approx
\frac{1}{\sqrt{\eta}} \langle {\bf p}_c|\psi \rangle_{\bf 0} \equiv 
\frac{1}{\sqrt{\eta}} \psi({\bf p}_c),
\end{equation}
where ${\bf p}_c$'s components perpendicular to ${\bf q}$ are the same
as ${\bf p}$'s, but the component parallel to ${\bf q}$ is
${\bf p} \cdot \hat{q}/\eta$.  Keeping only this length-contraction
effect, at present, and approximating $\bar{u} \gamma^0 u$ by the first term
in its $p/M$ expansion---$1$---it is easy to show that:
\begin{equation}
{}_{{\bf q}/2}\langle m_z|J^0|m_z \rangle_{-{\bf q}/2}= F_1(Q^2) \int \frac{d^3
p_c}{(2 \pi)^3} \psi_M^*\left({\bf p}_c + \frac{{\bf q}}{2
\eta}\right) \psi_M({\bf p}_c).
\end{equation}

This is exactly the result for this matrix element in the
non-relativistic impulse approximation (NRIA), but with ${\bf q}$ replaced by
${\bf q}/\eta$. Expanding $\eta$ in powers of $Q/M$ one might think that the
difference between the NRIA and the instant-form RIA will be very small
unless $Q^2$ is large. In fact though, this effect is non-negligible,
especially near the minimum of $G_C$ at $Q \sim 800$ MeV, since the
charge form factor varies rapidly there.

\subsection{Survey of relativistic approaches}

This then is one relativistic effect, in one approach to relativistic
dynamics. What of other approaches?  A detailed description of various
models which attempt to deal with this regime can be found in the
recent reviews~\cite{GvO01,GG02,Si02}. I will provide no more than a
rough sketch here, in each case discussing only the relativistic
impulse approximation for each model. My plots for these results are
taken from the Gross and Gilman review~\cite{GG02}.

Considering calculations where quantum mechanical wave functions
generated in the rest frame are boosted to the Breit frame, we can
choose instant form, front form, or point form. Examples of all three
appear in Fig.~\ref{fig-compare}.  For the instant form we have a
calculation by Forest, Schiavilla, and Riska (FSR)~\cite{SP03}. It
proceeds essentially as described in the previous subsection, although
no $p/M$ expansion is made for the nucleon current, and Wigner
rotations are included (approximately) in the boost.  The input wave
function used is the AV18. This result includes some
two-body-current contributions.

For the front form we have two representatives: a calculation by Lev,
Pac\'e, and Salme (LPS), which works in a frame where $Q_\perp=0$.  In
Ref.~\cite{LPS} several input wave functions were used, but results
displayed here are for the NijmII wave function. We also show a
light-front calculation by Carbonell and Karmanov (CK), in which an
attempt to restore manifest rotational invariance is made~\cite{CK}.
This is done by allowing the orientation of the light front to be an
additional vector in the theory, and then eliminating dependence on
it. Unfortunately, only approximate calculations in this
approach exist as yet.


Moving to the point form, I first note that for Eq.~(\ref{eq:sing}) to
be correct I should have $j_\mu$ depending on ${\bf k}_1'-{\bf k}_1$.
In the point form (and also in some versions of the front
form~\cite{LPS}) this need not be the same as ${\bf q}$~\cite{AKP}. A
consequence of this is that the squared momentum transfer to the
struck nucleon is $f({\bf p}) Q^2 \eta^2$, where the function $f \gsim
1$, but depends on the nucleon relative momentum ${\bf p}$. A calculation by
Allen, Klink, and Polyzou (AKP) using this form of dynamics, together
with the AV18 wave function is displayed below~\cite{AKP}

In any one of these forms of quantum mechanics one can take the
one-body current operator and evaluate its matrix element between
deuteron wave functions, defining that as the relativistic impulse
approximation. However, in the front form with $Q^+=0$ (and also in
the point-form calculation of AKP) an impulse approximation such as
this implicitly includes some two-body current effects.  That is
because the full consequences of Poincar\'e covariance and current
conservation are not necessarily respected in calculations of
individual matrix elements, and must be imposed on the calculation by
a procedure equivalent to introducing two-body currents which act
specifically to restore these symmetries.

Also shown in Fig.~\ref{fig-compare} are two calculations in which
$|\psi \rangle_{\pm {\bf q}/2}$ is calculated dynamically, that is to
say, the dynamical equation of the theory in the moving frame is
solved to obtain $|\psi \rangle$. Obviously in order to do this one
needs an $NN$ model in which $V$ can be calculated in an arbitrary
frame, and this is simplest for meson-exchange models. The calculation
of van Orden, Devine and Gross~\cite{vO95} (vOG) uses an interaction
developed in the ``spectator formalism'' and fit to $NN$
data~\cite{Gr92}.  That of Phillips, Wallace, and Devine (PWD) employs a
two-body equation which incorporates the effects of Z-graphs and
relativistic kinematics in the $NN$ system, but only has approximate
boost covariance~\cite{PWD}. There the Bonn-B interaction is
chosen. Both calculations have impulse approximations which respect
deuteron four-current conservation.

\subsection{Results}

\begin{figure}[htbp]
\centerline{\epsfig{figure=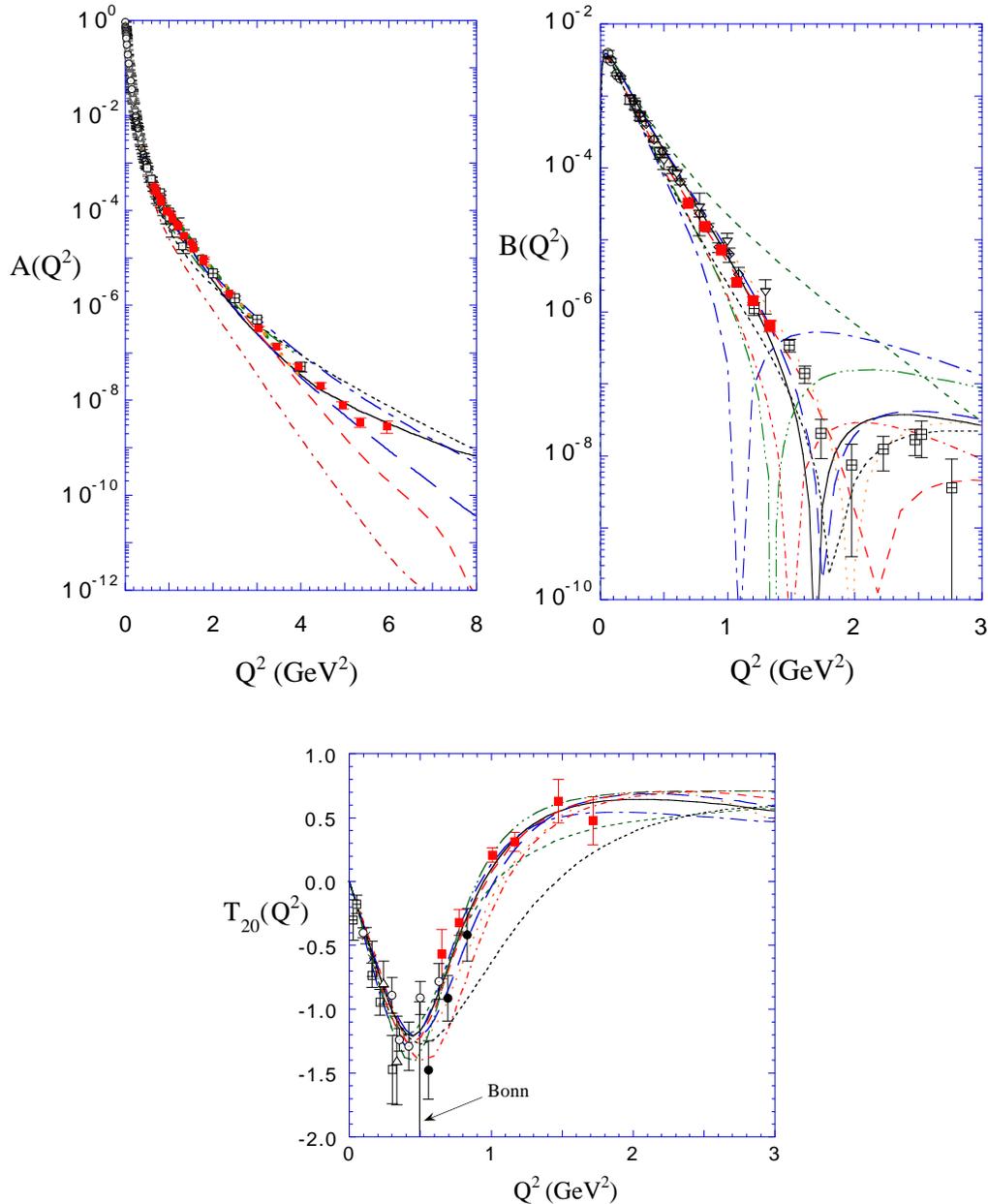, width=14cm}}
\caption{Comparison of different relativistic calculations for $A$,
$B$, and $T_{20}$. All calculations use single-nucleon form factors
due to Mergell {\it et al.}~\cite{Me95}. The calculations, in
order of their minima in $B$, are: CK (long dot-dashed line), PWD
(dashed double-dotted line), AKP (short dot-dashed line), two versions
of the vOG calculation (solid line and long-dashed line), LPS (dotted
line), a quark-model calculation not discussed here (widely-spaced
dotted line), FSR (medium-dashed line), and a calculation using a
$p/M$ expansion by Arenhoevel, Ritz, and Wilbois~\cite{Ar98}, also not
discussed here (short-dashed line).  Figure from Ref.~\cite{GG02},
thanks to Ron Gilman and Franz Gross.}
\label{fig-compare}
\end{figure}

Turning now to the results shown in Fig.~\ref{fig-compare} we see that
there are a wide range of RIA predictions. All calculations predict a
rapid fall in $A$, but they agree with the experimental data to
differing degrees, especially once $Q^2 \sim 2$ GeV$^2$.  Agreement
with the data at the level of the RIA is not essential for success,
since explicit two-body currents can, and probably should, be added to
all these calculations. However, large two-body currents might cast
doubt upon the efficiency of an expansion in hadronic degrees of
freedom. For further discrimination between models it is good to
divide out the rapid overall trend, as is done in Ref.~\cite{GG02}. I
will not do a detailed comparison of models with experiment here.

What is clear is that even were the data to be removed from the slide,
there would be no definite prediction for the RIA value of $A$ at,
say, $Q^2=2$ GeV$^2$. This is true even though most of these calculations
respect the Poincar\'e algebra and current conservation, and remains
true even if we restrict ourselves to those calculations using the
same rest-frame dynamics (e.g.~AKP and FSR). Symmetries and rest-frame
dynamics alone are not enough to unambiguously predict $A$ at these
high values of $Q^2$.

This is perhaps even clearer in $B$, where the position of the minimum is
an extremely sensitive test of the dynamics: both the implementation
of the Poincar\'e algebra and the choice of potential. But again, it
is worth stressing that even calculations with the same rest-frame
dynamics produce markedly different predictions for the minimum:
the position of the minimum in the AKP and FSR calculations differs
by about 30\%. And front form and point form even predict opposite
directions of the ``relativistic shift'' of this minimum~\cite{LPS,AKP,CP}.


Finally, in $T_{20}$, most models seem to do reasonably well.  The
JLab Hall C experiment~\cite{Ab00A}, which extended measurements of
$t_{20}$ out to $Q^2=1.7$ GeV$^2$, shows that relativistic hadronic
models of deuterium have some success in predicting this observable.
However, detailed comparison with the accurate data at $Q^2 \lsim
0.5$ GeV$^2$ reveals that models have some problems in precisely
reproducing this data.


To further understand the range of predictions for $A$ and $B$ above 1
GeV$^2$ I want to consider a gedankencalculation, or actually two
gedankencalculations.  Calculations 1 and 2 both begin with the same
rest-frame $|\psi \rangle$ and the same one-body $j^\mu$, but they have
two different implementations of the Poincar\'e algebra, i.e.
\begin{equation}
|\psi \rangle_{\bf Q}^{(1)}=U_{\bf Q}^{(1)} |\psi \rangle_{\bf 0}; 
\qquad  |\psi \rangle_{\bf Q}^{(2)}=U_{\bf Q}^{(2)} |\psi \rangle_{\bf 0}.
\end{equation}
Since the two $U$'s are both unitary representations of the Poincar\'e algebra
there is a unitary transformation that can make
an exact equivalence between matrix elements in formulation 1 and
those in formulation 2. If I choose:
\begin{equation}
J^{(2)}_\mu=U^{(2)}_{{\bf q}/2} {U^{(1)}_{{\bf q}/2}}^\dagger
j_\mu U^{(1)}_{-{\bf q}/2} {U^{(2)}_{-{\bf q}/2}}^\dagger,
\label{eq:unit}
\end{equation}
then:
\begin{equation}
{}^{(2)}_{{\bf q}/2}\langle \psi|J^{(2)}_\mu|\psi \rangle^{(2)}_{-{\bf q}/2}
={}^{(1)}_{{\bf q}/2}\langle \psi|j_\mu|\psi \rangle^{(1)}_{-{\bf q}/2}.
\label{eq:unitequiv}
\end{equation}
Crucially though, the unitary transformation (\ref{eq:unit}) results
in $J_\mu^{(2)}$ having two-body pieces, even though $j_\mu$ is only a
one-body operator. Thus, the equivalence (\ref{eq:unitequiv}) only holds
if these two-body currents are included in $J_\mu^{(2)}$. If the relativistic
impulse approximation is used in formulation 2 also, then
the results of the two different calculations will differ, by
an amount given by the size of the two-body piece induced
in Eq.~(\ref{eq:unit}). 

\section{Conclusion}

\label{sec-conc}

Henri Poincar\'e once said ``Les faits ne parlent pas'': facts do not
speak.  Nevertheless, let me attempt to interpret the results presented
here.

I draw two lessons from them. Firstly, {\it there is no unique
relativistic impulse approximation prediction for electron-deuteron
scattering}. The RIA is only defined once a particular implementation
of the Poincar\'e algebra is chosen. Associatedly, the size of the
variation in Fig.~\ref{fig-compare} is indicative of the size of
two-body contributions to $A$ and $B$.

Does this mean then, that the situation is hopeless? Not at all: it is
just that when looking at an RIA result one must know how the
Poincar\'e algebra was implemented. It also suggests that in any given
formalism it is important to examine two-body currents. Different
formalisms will have different two-body currents, and it would seem
that in some approaches these currents are quite sizeable. Work in
this direction has already been done in instant form~\cite{SP03}, and
is progressing in front form~\cite{CP}, and in a calculation 
involving dynamical wave functions~\cite{PW}.

More data would be useful in order to help different theorists pin
down these contributions in their formalisms. A theory-independent
statement about the size of two-body currents is impossible at $Q^2
\sim 2$ GeV$^2$, but once a specific theory is chosen data is needed
to calibrate the two-body currents which contribute significantly to
elastic electron-deuteron scattering at these momentum
transfers. Especially useful would be more data around the minimum of
$B$, since this is a barometer sensitive to many model elements.

Having worked hard to tune-up our calculations to reproduce elastic
electron-deuteron data, an important future step is the extension of
these calculations to deuteron electrodisintegration (for steps in
this direction see~\cite{SR91,Ad02}).  Finally, much of what was said
here also applies to the three-nucleon system. The boosted $NN$
t-matrix is input to the Faddeev equations for $NNN$ bound and
scattering states.  Different implementations of the Poincar\'e
algebra lead to different boosted $NN$ t-matrices. The unitary
equivalence between these implementations is maintained only if the
associated three-body forces are kept in the three-nucleon calculation.

\section*{Acknowledgements}
I thank Fritz Coester, Ron Gilman, Franz Gross, Brad Keister, Jerry
Miller, Wayne Polyzou, John Tjon, and especially Steve Wallace for
stimulating discussions on the topics discussed herein and, in some
cases, for sharing their results with me for use in my talk. This work
was supported by the U.~S. Department of Energy under grants
DE-FG02-93ER40756 and DE-FG02-02ER41218.

\end{document}